# Business Processes of High-Tech Enterprises

S.E. Pyatovsky

*PhD in Economic sciences, Associate Professor of the Department of «Management of high-tech enterprises», Moscow Aviation Institute (National Research University)
(e-mail:vgsep@ya.ru)*

**Abstract.** This paper analyzes the results of Russia's current innovative activities. It shows the need to increase the level of return on investments in the innovative capacities of high-technology enterprises (HTEs). The paper describes the methods to help increase the competitiveness of HTEs based on the implementation of modern control methods for innovative HTEs. It analyzes HTE KPIs, and also describes the characteristics of the organizational structure of HTEs. HTEs are studied as an innovative self-training HTE, and their characteristics and system of competencies are analyzed. The paper likewise describes the management of information support for managerial decisions in self-training HTEs. It shows that a considerable share of innovative products in a highly competitive market requires accelerated promotion and new approaches to building the system of business processes (BPs) for self-training HTEs. It also shows the dependency of self-training HTE competitiveness on innovative management methods for manufacturing and technological processes (MTP).

## Introduction

Organizational planning of the HTE includes BPs, management systems and automated economic information systems (AEIS) that determine the quality of manufactured products, expenses, prices and communication channels. HTEs provide on-time product supplies to end consumers, which implies a high level of standardization and automation in the applicable BPs. HTE management systems are focused on operational efficiency, organizational planning, - on minimization of expenses.

The main areas of focus to decrease HTE [1] product costs include the minimization of logistics expenses, the centralized planning of standardized products, and the implementation of high-level, standardized, integrated automation systems of technological processes management (ASTPM). The implementation of projects using AEIS is required to increase HTE competitiveness.

The subject of scholarly discussions on this issue revolves around how to manage HTE BPs. Industry based on high technology forms the core of developed economies. Currently, the innovative activity of Russia is in a difficult state. Table 1 shows the Global Innovation Index (GII) [2] of Russia and the USA for 2011–2017. The GII shows positive dynamics with a growth rate of 0.77 per year ($R^2 = 83\%$) (USA) and 0.43 per year ($R^2 = 57\%$) (Russia). The innovative activities of Russia are improving despite the relatively high volatility ($R^2 = 57\%$) caused by systemic risks. At the same time, GII components show an inverse ratio of innovative growth (Figure 1).

Table 1
GII of Russia and the USA (2011–2017) [2]

| Year | 2011 | 2012 | 2013 | 2014 | 2015 | 2016 | 2017 |
|---|---|---|---|---|---|---|---|
| Russian Federation | 35.85 | 37.90 | 37.20 | 39.14 | 39.33 | 38.50 | 38.76 |
| United States of America | 56.57 | 57.70 | 60.31 | 60.10 | 60.10 | 61.40 | 61.40 |

Figure 1 presents the dynamics of GII Input/Output [2] of Russia and the USA for 2011–2017. The GII Input Sub-Index characterizes the conditions and factors necessary to create innovations, and includes such indicator groups as institutions, human capital and research, infrastructure, market and business stability. The GII Output Sub-Index characterizes the results of innovative activities, and includes such indicator groups as scientific and creative results, and a creative online approach. The initial GII indices are determined by min-max normalization and characterize the government's innovative activities.

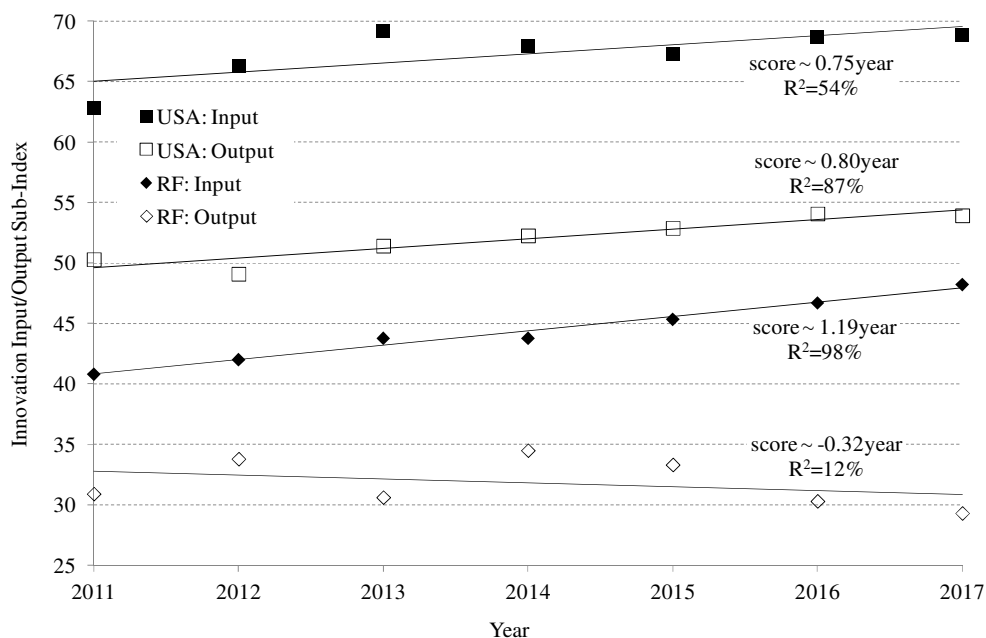

Figure 1. Dynamics of innovative activities of Russia and the USA for 2011–2017
(GII methodology [2])

It follows from Figure 1 that the dynamics of innovative activities in Russia are considerably higher than similar dynamics in the USA. For the last 6 years, the growth in the GII Input Sub-Index for the USA has been 0.75 per year ($R^2 = 54\%$), and the growth in the GII Input Sub-Index for Russia has been 1.19 per year ($R^2 = 98\%$). But the results of innovative activities in Russia and the USA show an inverse ratio of indices. The growth in the GII Output Sub-Index for the USA is 0.80 per year, $R^2 = 87\%$, which complies with the GII Input Sub-Index (0.75 per year, $R^2 = 54\%$). The growth in the GII Output Sub-Index for Russia shows negative growth of -0.32 per year and a high volatility ($R^2 = 12\%$). This implies that Russia regulates investments in innovative institutions, but the return on investment has a random character determined by high systemic risks.

An increase in the government's return on investment in research and development (R&D) is determined by the increasing competitiveness of the HTE, which is an indicator of the economic development of the country. The current situation requires new approaches from HTEs to the AEIS and ASTPM arrangement. The importance of innovative projects is considered in scientific publications [3,4], and by the management of leading international corporations [5,6].

HTEs analyze data on sales and operations to maximize the efficiency of organizational planning, minimize expenses and increase market share [7]. HTEs engaged in complex manufacturing analyze the information of managerial decision support with the use of the AEIS. HTEs that quickly form a knowledge database in the information circuit receive a competitive advantage and are better prepared for managerial decision making. To achieve the maximum efficiency of the AEIS, the system should feature high-performance and cheap information analysis tools [8].

Scientific publications [9] show that the solution of these tasks requires the algorithmization of modeling of the HTE BPs system as an element of the HTE business strategy based on competencies. Building a hierarchical model of management for HTE information resources at the instrumental, object and procedural levels [10] is a necessary condition for HTE operations in a highly competitive market.

HTE information resources provide for the implementation of R&D programs. Scientific publications [11] propose solutions to improve the management efficiency of HTE resources, including the processing of information of managerial decision support using dynamic programming methods. It is shown that HTEs operates in the conditions of an innovative economy rapidly brings R&D solutions to the market using modern IT [12].

1. Organizational Planning of HTEs

Organizational planning of HTEs is focused on the analysis of production, factors of the internal and external environment, strategic analysis, and definition of goals and alternatives. Organizational decisions are reflected as an operational plan containing the budgets and general production schedule.

Organizational planning is represented by operational and strategic planning on the basis of detailed information from the corporate information system (CIS) [9,13]. The task solving efficiency for the optimization of R&D programs of HTE by dynamic programming methods is shown in [11], where the decision process is divided into $j$ time of $S$ duration $= T_{pl} / j$, where $T_{pl}$ is the period of the R&D program with a direct time account. By specifying with $\{N_k\}$ a set of work programs in the $k$ optimization step, the mathematical model of the task is defined by:

$$F = min \sum_{k=1}^{j} f_k[N_k + \psi(\overline{P}, N_k)], \qquad (1)$$

where:

$f_k(N_k)$ is a function of total losses determined by the shift of the program completion time $\{N_k\}$ in the $k$ optimization step; $\psi(\overline{P}, N_k)$ is the function of additional expenses and losses determined by the deviation of resources used by $\{N_k\}$ from the specified value of HTE resources available $\overline{P}$ in the $k$ optimization step.

The functional equation of the task solution (1) results from the recurrence relation:

$$f_k(\overline{\varepsilon}_k) = min\{f_k(N_k) + \psi(\overline{P}, N_k) + f_{k+1}[\varphi_{k+1}(L)]\}, \qquad (2)$$

where: $\{L\}$ is the set of R&D programs that pass from the $k$ to $k+1$ optimization step; $\varphi_{k+1}(L)$ are the resources that pass from the $k$ to $k+1$ optimization step; $f_k(\overline{\varepsilon}_k)$ is the evaluation function of $i$-$k$+1 step process the initial state of which is $\overline{\varepsilon}_k$; $f_{k+1}[\varphi_{k+1}(L)]$ is the evaluation function of $i$-$k$ step process the initial state of which is $\varphi_{k+1}(L) = \overline{\varepsilon}_{k+1}$.

Relation (2) determines the optimal value of the HTE target function when completing the $i$-$k$+1 optimization steps for the $\overline{\varepsilon}_k$ vector and characterizes the initial state of the system during the $k$ optimization step. The $\{\overline{\varepsilon}_k = 0, \overline{\Delta}, \ldots, \overline{P}_{max}\}$ set is a set of states, in one of which the system can be in the $k$ step with the optimal number of works in the $k$ step $\{N_k\}$. The results of calculations for equations (1) and (2) is the matrix of planning strategies of $N_k(\varepsilon_k)$ for $\overline{\varepsilon}_k = \overline{\varepsilon}_{k+1} = 0$.

HTEs consist of strategic business units (SBU). SBUs meet the requirements of the availability of enterprise resources planning (ERP), individual strategies, the strategic plan, and unique organizational advantages. The management of the HTE SBU has coordination, management and control functions.

The factors determining HTE behavior include a small number of buyers, stable relationships between buyers and suppliers, the geographical concentration of consumers, derivative and inelastic demand with high volatility, and a significant number of decision makers. To address these issues, HTEs create a management decision center (MDC). The MDC in the circuit of HTE CIS [7,14,15] takes into account that services are offered to customers at all stages of the procurement process. With an increase in the risk connected with a new supply, the procurement process becomes more complicated, which increases the information of managerial decision support formed by the AEIS.

The HTE procurement process controlled by the AEIS includes the establishment of functional requirements and product specifications, the development of criteria for measuring MTP effectiveness, synchronization of external and internal information, determination of proposal volumes, formation of the pricing policy and conditions of deliveries of products, making decisions on suppliers and building feedback through HTE communication channels.

HTE organizational planning provides HTE leadership on the product, which means the high-speed introduction of innovative R&D results and MTP improvement. To implement this strategy, the high customer-oriented approach of HTEs with a focus on R&D, market expansion, innovative BPs and the unified information space of the HTE are required.

Leading HTEs in terms of products offer innovations in accordance with strategic priorities, which forms the system of intangible assets of the HTE, including the brand and goodwill.

**2. Strategic Planning in HTEs**

Strategic planning in HTEs includes SBU profile analysis, confrontational analysis, strategic break point analysis, and the adjustment of strategic alternative profiles.

Due to the significant role HTEs play in the economy [16,17], the R&D directions of HTEs include attention spent [18] on determining product functions, the target audience profile, the MTP and the level of the SBU trading chain. Strategic breaks in R&D planning depend on the strategic attention points of the HTE. It has been shown [1] that the internal rate of return on R&D investment project of HTEs representing the *d* discount norm is determined by:

$$\sum_{i=0}^{T} \frac{R_i}{(1+d)^i} = \sum_{i=0}^{T} \frac{K_i}{(1+d)^i} \qquad (3)$$

It follows from relation (3) that *d* is determined by the equality condition of the given *R* effects and *K* capital investments. The discount of an R&D project determines the upper limit of the acceptable income rate for the capital invested in the project.

The analysis of strategic breaks is carried out for all supply and marketing activities of the HTE. The analysis of strategic break points shows the consequences of organizational planning, but not the causes of the break. A significant number of tools have been developed to reduce strategic breaks, such as the development and implementation of HTE adapted growth strategies.

The analysis of micro and macro factors is followed by HTE strategic alternatives [19]. The use of innovative technologies of the receipt and processing of information for managerial decision support provides for HTE efficiency in a highly competitive market [20]. The modelling of economic systems is necessary to control HTE innovative development and includes the efficiency estimation for the use of the innovative potential of the HTE and manufactured products. The latter contributes to improving the efficiency of the innovative modernization of the country's industrial complex. It is shown [12] that the long-term planning of alternatives offers a solution for a key HTE issue and corresponds to the HTE business environment and SMART criteria.

To evaluate alternatives, information on product test samples and investment analysis is required. But even with complete market information, it is impossible to forecast all alternatives. The solution is to draw up several forecasts according to scenarios of the positive, negative and neutral development of the market environment. The process of receiving managerial decision support information for HTEs is characterized by the fact that the behavior of the buying enterprise is rational, wherever the theory of the "black box" is not applicable.

### 3. Efficiency Factors of HTEs

Scientific publications [21] show the efficiency of HTE KPI groups:

- KPIs that characterize break-even points, order delivery terms, number of complaints, the qualification of engineering and technical staff (ETS), and equipment in specialized educational institutions;

- KPIs that characterize the motivation of employees and customers, innovation in production, mobility of the HTE, a single information space, and the efficiency of the HTE.

HTE efficiency when forming a portfolio of strategic alternatives is determined by the system of equations:

$$\begin{cases} \min \max[\Psi_i(\cdot)] \\ \text{System of external confines} \\ \text{System of internal confines} \\ \text{System of normative confines} \end{cases} \quad (4)$$

where $\min \max[\Psi_i(\cdot)]$ is a min-max criterion for selecting the $i$ target function of an alternative R&D project.

During the implementation of criteria (4), the HTE system is stable in the cyber definition. The target functions in the dynamic model of BPs help form short, medium and long-term forecasts. The optimization task has a lot of criteria, and requires the formation of a combined optimization criterion. The basic method for building a combined criterion is the goal programming method where the optimal criterion $\Psi^0 = \{\Psi_i^0\}$ is determined by the indicators of the leading international corporations, with the solution of optimization tasks:

$$\min ||\Psi, \Psi^0|| = \min \left( \sum_{i=1}^{n} a_i \, |\Psi_i(\cdot) - \Psi_i^0(\cdot)|^n \right)^{1/n}, \quad (5)$$

where $a_i$ is the weight coefficients determining the criterion significance.

The KPI combination in equations (4) and (5) is the basis of the key competencies of the HTE. These KPIs include the level of staff qualifications, and the recruitment and retention of staff, including the definition of standards and goals. In a highly competitive environment, HTE KPIs are determined by rapidly changing market needs. The improvement of HTE competitiveness is provided by the KPI of product and process innovations, the level of diversification of production activities, entry into the international market as a supplier of components, and the use of modern financial and economic instruments and schemes.

**4. Innovative Self-Training HTEs**

HTEs implement a matrix organizational structure allowing to use ETS for various R&D projects [10]. This form of organization is applied in HTEs with R&D on a project basis. The ETS and resources are allocated for each project separately with the use of ASTPMs. Project team members are determined according to their competencies.

The next stage of HTE development is a self-training HTE with principles of collective leadership to maximize resource opportunities. Self-training HTEs offer continuous training and are constantly transforming. The system of competences for self-training HTEs is based on partnership relations. The process of improving the qualification of HTE ETS is performed at the personal and team levels, as well as at the level of partners the HTE interacts with. The HTE's investment in ETS is a strategic process integrated into R&D [11]. The result of training is reflected in the system of competences of the self-training HTE and expands HTE opportunities in the field of innovative development, which is a necessary condition for HTE success in a highly competitive market.

The system of collection, systematization, storage and use of knowledge is built into the information space of HTEs through the CIS. The process of ETS training is carried out at the individual and team levels, as well as at the BP and system levels. All systems, BPs and organizational structures search information for managerial decision support to provide self-training HTEs with competitive advantages and position R&D results. The key characteristic of self-training HTEs is their search and use of feedback for strategic development.

The characteristics of self-training HTEs are considered at the system level and at the levels of subsystems and BPs [7,22]. The characteristics of self-training HTEs at the system level include the ability of self-training HTEs to consider clients as partners, the implementation of the organizational planning of self-training HTEs in terms of feedback from the external and internal environment, the maximization of the number of employees who have the opportunity to improve their skills, and an employee motivation scheme focused on problem solving.

The characteristics of self-training HTEs at the levels of subsystems and BPs include feedback from the external and internal environment of the HTE, the improvement of ETS skills in accordance with the HTE mission, and the ETS motivation system.

This requires the creation of a single information space within the HTE [7,10,23] for work with multidimensional database structures to store consolidated information for managerial decision support. Due to the specific character of HTE operations, the single information space requires a hierarchical system of access rights differentiation at the tool, object and procedure levels. At the tool level, users are given the opportunity to work with a set of ASTPM tools determined by the structure of the BP control system; at the object level, users are given the opportunity to work with a set of information objects; at the procedure level, users are given different rights when working with database information objects.

**Conclusions**

The study of the organizational planning features of HTEs in a highly competitive market showed the need to apply innovative approaches to the arrangement of BPs. Innovations are studied in relation to the MTP, the distribution system and the management BPs [12]. The specific features of innovative BP products are provided with the characteristics of the R&D MTP, including the idea search and its development to product concept, pilot production, product introduction on the experimental market, and the commercialization process.

Innovative products occupy a significant share of a highly competitive market. The knowledge economy requires accelerated introduction on the market and modern approaches to the modeling of the BPs system. New methods of HTE organizational planning, including developed system of competencies, are a necessary condition for the strategic competitive advantage of HTE.

*Bibliography:*

1. Pyatovsky S. E. Application of financial marketing tools in the evaluation of investments in information technology / S. E. Pyatovsky // Marketing and finance. 2013. № 2: 124–130. URL: https://goo.gl/Hajof1 (accessed 15.04.2018).
2. The Global Innovation Index // The Business School for the World. URL: https://www.globalinnovationindex.org
3. Daily J. Predictive Maintenance: How Big Data Analysis Can Improve Maintenance / J. Daily, J. Peterson // Supply Chain Integration Challenges in Commercial Aerospace. Springer International Publishing, 2017: 267–278. DOI: http://dx.doi.org/10.1007/978-3-319-46155-7_18
4. Komarov V. A. Intellectual data analysis in aircraft design / V. A. Komarov, S. A. Piyavskiy // CEUR Workshop Proceedings. 2016. Vol. 1638: 873–881. DOI: http://dx.doi.org/10.18287/1613-0073-2016-1638-873-881

5. High P. Interview with T. Colbert: Boeing CIO Ted Colbert drives digital and security transformation / P. High // Forbes. 2016. March 14. URL: https://goo.gl/xbXiWx (accessed 15.04.2018).

6. Rajpurohit A. Interview with D. Kasik: Boeing on Data Analysis vs Data Analytics / A. Rajpurohit // Kdnuggets, 2015. Feb. URL: https://goo.gl/Baf7QY (accessed 15.04.2018).

7. Pyatovsky S. E. Manufacturing processes hardware and software development to implement innovative technologies of aircraft manufacturing facilities management / S. E. Pyatovsky, A. N. Serdyuchenko // Journal of Economy and entrepreneurship. 2018. Vol. 12. № 1: 856–859. URL: https://goo.gl/ZzZTaa (accessed 15.04.2018).

8. Yost K. System and method for management of an automatic OLAP report broadcast system / K. Yost, and others // pat. 9477740 USA. 2016. URL: https://goo.gl/5zVfBu (accessed 15.04.2018).

9. Pyatovsky S. E. Algorithmization of business processes as an instrument of managing marketing assets / S. E. Pyatovsky, A. V. Degtyarev // Herald of Moscow State Linguistic University (MSLU). Ser.: Economics. The problems of modernizing the Russian economy at the present stage. 2010. Issue 6 (585): 55–66. URL: https://goo.gl/aH6iek (accessed 15.04.2018).

10. Pyatovsky S. E. The model of distinction of access rights to information objects of the system of controlling of business processes of an aviation enterprise / S. E. Pyatovsky, A. V. Degtyarev, V. A. Vdovin // Statistics and Economics: Plekhanov Russian University of Economics (Moscow). 2014. №3: 38–44. URL: https://goo.gl/apAFwr (accessed 15.04.2018).

11. Pyatovsky S. E. The optimization of a resource component of an aviation enterprise's investment portfolio / S. E. Pyatovsky, V. A. Oganov // The successes of modern science. 2016. V.1. №6: 16–19. URL: https://goo.gl/hpc3qo (accessed 15.04.2018).

12. Pyatovsky S. E. The concept of the automated project management system for an Aviation Enterprise / S. E. Pyatovsky // Journal of Economy and entrepreneurship. 2017. №5 (2): 803–806. URL: https://goo.gl/oW7uBV (accessed 15.04.2018).

13. GOST R ISO/IEC 12207-2010 «Information technology. System and software engineering. Software life cycle processes». URL: https://goo.gl/ct58aC (accessed 15.04.2018).

14. Pyatovsky S. E. Fundamentals of the algorithmization and modeling of business-processes / S. E. Pyatovsky, A. V. Degtyarev // M.: MAI. 2009.

15. Back W. D. Mondrian in action. Open source business analytics / W. D. Back, N. Goodman, J. Hyde // 2014. ISBN-13: 978-1617290985, ISBN-10: 161729098X

16. Bourguignon F. Air traffic and economic growth: the case of developing countries / F. Bourguignon, P. E. Darpeix // Sciences de l'Homme et Societe. Economies et finances, 2016. URL: https://goo.gl/wVKJVJ (accessed 15.04.2018).

17. Mehmood B. Air transport and macroeconomic performance in Asian countries: An Analysis / B. Mehmood // Pakistan Journal of Applied Economics. 2015. Vol. 25. № 2: 179–192. URL: https://goo.gl/vXkY1v (accessed 15.04.2018).

18. Pyatovsky S. E. Formalization and modeling of business processes (structural approach) / S. E. Pyatovsky, A. V. Degtyarev // M.: MAI. 2010.

19. Pyatovsky S. E. Algorithmization of business processes in goal to identify strategic points of attention of the company / S. E. Pyatovsky // Actual marketing technologies in the development of the Russian economy: Financial University under the Government of the Russian Federation. 2012. 292 p. ISBN 978-5-94727-629-9.

20. Maranville S. Entrepreneurship in the Business Curriculum / S. Maranville // Journal of Education for Business. 1992. Vol. 68. № 1: 27–31. DOI: http://dx.doi.org/10.1080/08832323.1992.10117582

21. Pyatovsky S. E. The development of the marketing strategies of an organization based the analyze of the financial markets / S. E. Pyatovsky // Herald of the Moscow State Linguistic University. 2014. Issue 6 (692): 119–130. URL: https://goo.gl/B54HFM (accessed 15.04.2018).

22. Pyatovsky S. E. Technology for administering of the access to information resources in management system on the aviation enterprise / S. E. Pyatovsky, A. V. Degtyarev, V. A. Vdovin // Statistics and Economics: Russian University of Economics (Moscow). 2015. № 1: 188–193. URL: https://goo.gl/mwQ8Dk (accessed 15.04.2018).

23. Kuznetsov S. D. Data management: 25 years of Forecasts / S. D. Kuznetsov // Proceedings of the Institute of System Programming, RAS. 2017. Vol. 29. Ser. 2: 117–160. DOI: http://dx.doi.org/10.15514/ISPRAS-2017-29(2)-5